\documentclass[%
 reprint,
 amsmath,amssymb,
 aps,
]{revtex4-2}
\makeatletter
\def\@biblabel#1{}
\makeatother

\usepackage{cancel}
\usepackage{graphicx}% Include figure files
\usepackage{dcolumn}% Align table columns on decimal point
\usepackage[normalem]{ulem}
\usepackage{amsmath}
\usepackage{amssymb}
\usepackage{color}
\usepackage[utf8]{inputenc}
\usepackage{graphicx}
\usepackage{bm}
\usepackage{mathrsfs} % Skri symbols
\usepackage{braket}
\usepackage{hyperref}
\usepackage{tikz}
\usepackage{bm}% bold math
\begin{document}

\preprint{APS/123-QED}

\title{Deriving the
Gibbons-Maldacena-Nunez  no-go theorem from the Raychaudhuri equation}% Force line breaks with \\

\author{Mir Mehedi Faruk}
\affiliation{%
 Department of Physics, McGill University, Montreal, Quebec, H3A 2T8, Canada.
}%

\date{\today}% It is always \today, today,

\begin{abstract}In this article, we point out that to solve the null Raychaudhuri equation for higher dimensional spacetime with accelerating FRW solution in external directions and static compact internal directions, it is necessary to violate the Strong Energy condition in higher dimensions. This constraint is well-known in obtaining accelerating cosmological solutions in string compactification, first described by Gibbons-Maldacena-Nunez. In deriving this constraint, we do not make any assumptions regarding the matter content.
\end{abstract}

\maketitle

%\tableofcontents
We have compelling evidence that the present universe is dominated by dark
energy and going through a period of accelerated expansion.
It is, therefore, natural to ask ourselves what quantum gravity theories, such as string/M-theory
can tell about accelerating cosmological solutions\cite{Danielsson:2018ztv}.
However,
 the construction of accelerating cosmological solutions in string theory is still an open problem even after almost twenty years of the subject.
The research direction took an interesting turn
due to the obstacles
such as the no-go theorems\cite{Gibbons:2003gb,
Maldacena:2000mw} and the
proposed swampland conjectures\cite{Grana:2021zvf,Brennan:2017rbf,Obied:2018sgi}.
The most interesting of them is the well-known no-go first proposed by Gibbons\cite{Gibbons:1984kp,Gibbons:2003gb}; later on, a more refined version came from Maldacena and Nunez\cite{Maldacena:2000mw} which took into account the supergravity fluxes and the (Anti) D-Branes. The most significant point made in this no-go theorem is that to obtain a $d$-dimensional accelerating cosmological solution by compactifying a $D>d$ dimensional theory, it is necessary to violate the $D$-dimensional strong energy condition (SEC). Afterward, many other no-go theorems were constructed in string cosmology paradigm from different viewpoints, including metric-based constraints\cite{Koster:2011xg,Wesley:2008fg,Montefalcone:2020vlu,Steinhardt:2008nk},
worldsheet symmetry\cite{Parikh:2015wae,Parikh:2014mja,Parikh:2016lys},
energy conditions
\cite{Bernardo:2021zxo,Parikh:2015ret,Russo:2018akp,Russo:2019fnk,Koster:2011xg,Bernardo:2022ony,Obied:2018sgi,DeLuca:2021pej,Alexander:2023qym}, supersymmetry\cite{Basile:2020mpt,Anous:2014lia,Basile:2021mkd} string/M theory \cite{Kachru:2003aw,Dasgupta:2018rtp,Dasgupta:2014pma,DeLuca:2021pej,Kutasov:2015eba,Townsend:2003fx,Teo:2004hq,Chen:2003ij,Wohlfarth:2003ni,Fischler:2001yj,Roy:2003nd,Marconnet:2022fmx,Russo:2022pgo,Dasgupta:2018rtp,Dasgupta:2019gcd,Cornalba:2002fi,Dasgupta:2014pma,Ferrara:2019tmu,Kounnas:2013yda,Townsend:2004zp,Khoury:2001bz,Emelin:2020buq}, spacetime entropy and quantum gravity in dS space \cite{Klemm:2004mb,Witten:2001kn,
Goheer:2002vf,Aalsma:2022eru,Chapman:2022mqd}. We have made considerable progress in understanding the properties of gravitational thermodynamics
\cite{Jacobson:1995ab,Eling:2006aw,Kar:2006ms,Jacobson:2015hqa,Chirco:2009dc}. In these works as well as in the
Hawking-Penrose
singularity theorems\cite{Hawking:1973uf,Hawking:1970zqf}
Raychaudhuri equation provides a fundamental contribution. 
\\\\
Raychaudhuri equation, a 
well-known geometric identity
is
used extensively to enhance our understanding of various disciplines involving
gravity, from astrophysics\cite{DiPrisco:1994np} to holography\cite{Alvarez:2000jb} and quantum gravity\cite{
Das:2013oda, Burger:2018hpz}. Recently, in an interesting article\cite{bret} this geometric identity has been used to further derive new no-go theorems in string compactification.
In the no-go theorem, the article concluded that accelerating
backgrounds in string theory can only solve the Raychaudhuri equation when the null energy condition (NEC) is violated and/or the
 internal directions have positive curvature. Besides,
 the well studied flux  compactification schemes in dS, such as KKLT, are 
 re-visited in light of this new no-go theorem in the article\cite{bret}.
The authors point out that
 the matter sources or geometries that can potentially
evade many of the previous no-go's and are considered to be essential ingredients
in building putative dS solutions are unfortunately ruled out by the NEC violation constraint\footnote{Third paragraph of page 2 of ref. \cite{bret}}.
  This is bad news for de Sitter (dS) compactifications; as we know, four-dimensional dS
 maintains NEC\cite{Bire82}.
 The NEC is also well-known to be satisfied by a large set of matter content, which will further
restrict the models with extra dimensions.
In this article, we carefully study the conditions to satisfy the Raychaudhuri equation for a 
$D$-dimensional spacetime solution where we have a $d$-dimensional FRW solution in the external direction and a compact internal manifold of dimension $n$\cite{Giddings:2001yu}.
There are several arguments\cite{Danielsson:2018ztv} whether such cosmological background with accelerating cosmology can be obtained in string compactification.
The pursuit of constructing four-dimensional accelerating solutions in string theory has led to claims ranging from multiple solutions\cite{Kachru:2003aw,Cribiori:2019bfx,Linde:2020mdk} to none at all\cite{Obied:2018sgi,Ooguri:2018wrx}.
But here, we re-examine the constraints for such backgrounds to satisfy the Raychaudhuri equation.  In particular, we focus
if any violation of NEC is really necessary to satisfy the Raychaudhuri equation. It is well known that
NEC also plays a significant role
in establishing the existence of
the Big Bang singularity, as well as
proving the second law of thermodynamics for black holes\cite{Carroll:2004st}.
There are also some hints that
accelerating cosmology should
satisfy the NEC to have a
UV completion
in string theory\cite{Bernardo:2021zxo}.
For example,
the Virasoro constraint
in string theory
coming from worldsheet theory
 precisely gives rise to the NEC in the geometry\cite{Parikh:2014mja}. 
Because of NEC violation, it is quite difficult
to obtain wormhole solutions, 
 the creation of laboratory universes, and the
building of time machines\cite{Parikh:2014mja}.
 We demonstrate here that to solve the Raychaudhuri equation  with four-dimensional accelerating cosmology in  external direction successfully, we rather must violate the SEC in higher dimensions.
 It is quite remarkable as this is the precise statement of the GMN no-go theorem. 
We briefly discuss the energy conditions, the null Raychaudhuri equation, and the GMN no-go theorem. Then, in section IV, we work out the details to look for the constraints to solve the Raychaudhuri equation for the background metric in \eqref{x2}.
 
\section{A brief review on Energy conditions} 
We quickly discuss the Null and Strong energy conditions in this section and the constraints it provides on the  ''scale factor"  of the FRW metric[\onlinecite{Bire82}]. The NEC simply states that for all the nulllike vector $l^M$ we have the following constraint on Ricci tensor,
\begin{equation}
        R_{MN}(x) l^M l^N \geq 0,\quad  g_{MN}(x)l^M l^N = 0.
\end{equation}
Similarly,  SEC implies that for any timelike
vector $t^M$,
\begin{equation}
    R_{MN}t^M t^N\geq 0, \quad t^2 <0. 
\end{equation}
Considering the FRW metric in the physical time co-ordinates, 
\begin{equation}
    \tilde{g}_{\mu\nu}dx^\mu dx^\nu=-dt^2+a^2(t)\delta_{ij}dx^i dx^j
\label{xddddd2}.\end{equation}
As we know, $H=\frac{\dot{a}}{a}$ is the Hubble scale.
 Accelerating solutions are identified with 
 \begin{eqnarray}
 \Ddot{a}/a = \Dot{H} +H^2> 0.    
 \end{eqnarray}
 For a power law scale factor, $a(t) \propto t^{\gamma}$, we have
\begin{align}
    \text{SEC} &\iff 0<\gamma \leq 1,\\
    \text{NEC} &\iff \gamma\geq 0, 
\end{align}
 If NEC violation is necessary to maintain the Raychaudhuri equation,
 then it would rule out many the four-dimensional cosmology such as dS.   It is much more difficult to violate the
NEC than it is to violate the SEC.
 Violating the NEC,
on the other hand, is very difficult, and no known classical energy-momentum sources or
fields are known do so\footnote{We express our gratitude to the referee for assisting us in clarifying this issue.}.
 \section{The null Raychaudhuri equation}
In order to establish a singularity theorem, it is necessary to have an effective means of anticipating the appearance of focal points along  geodesics. Raychaudhuri's equation offers such a methodology\footnote{Interested readers please look into\cite{Witten:2019qhl}}. 
Raychaudhuri’s equation reveals that the occurrence of focal points is quite common, as gravity has a propensity to focus nearby geodesics. 
The Raychaudhuri equation for null geodesic congruences is the following, \begin{eqnarray} \frac{d\theta}{d\lambda} = -\frac{1}{D-2}\theta^2 -\sigma^2 - R_{MN}l^M l^N \label{ava} \end{eqnarray} Here, $l^N $ are the Null vectors. Now, the expansion parameter and the shear tensor are identified as, \begin{eqnarray} && \theta=\frac{1}{\sqrt{-g_D}}\partial_M(\sqrt{-g_D} l^M)\\ &&\sigma_{MN}=\frac{1}{2}(\nabla_M l_N+ \nabla_N l_M) -\frac{1}{D-2} \hat{h}_{MN}\theta \end{eqnarray} And, $\hat{h}_{MN}$ is the transverse metric, i.e., \begin{eqnarray} \hat{h}_{MN}l^M=0 \end{eqnarray} These are of course transverse to null rays.

\section{Gibbons-Maldacena-Nunez no go theorem}

We quickly review the GMN no-go theorem \cite{Gibbons:2003gb,Gibbons:1984kp,Maldacena:2000mw} in this section.
We will consider the following warped product $D=d+n$
dimensional manifold as a solution of higher dimensional
quantum gravity theory such as string theory ($D=10$) following\cite{Giddings:2001yu}. In the external direction, we took the four-dimensional FRW metric, and the internal direction is given by a (time-independent) compact $n$-dimensional manifold.
\begin{eqnarray}
ds^2=e^{-2A(y^m)}{g}_{\mu\nu}dx^\mu dx^\nu+
e^{2A(y^m)}{g}_{mn}(y^m)dy^mdy^n.\label{x101}\nonumber \\
\end{eqnarray}
The above metric can also be written in the following way using conformal rescaling of the internal metric\cite{bret},
\begin{eqnarray}
    ds^2=\Omega^2(y^m)[\tilde{g}_{\mu\nu}dx^\mu dx^\nu+
 \tilde{h}_{mn}(y^m)dy^mdy^n  ] \label{x2}\label{xGG}
 \nonumber \\
\end{eqnarray}
We will refer to the 
$\Omega$ as the warp factor. We take the compact
 manifold has no
boundary, and the warp factor is non-singular.
Calculating the Ricci tensor for the $D$-dimensional metric in the external directions following \eqref{bigchoppa},
\begin{eqnarray}    R^{(D)}_{\mu\nu}=R^{(d)}_{\mu\nu}(\tilde{g})
-\tilde{g}_{\mu\nu}[\nabla^2(\ln \Omega)+(D-2)(\nabla \ln \Omega)^2]\nonumber \\\label{chapoox}
\end{eqnarray}
The covariant derivative of the above equation is simply along the compact directions.
Using the Einstein equation for the full metric \eqref{xGG}, we find out that,
\begin{eqnarray}
    R^{(D)}_{\mu\nu}= T_{\mu\nu}-\frac{\Omega^2}{D-2}\tilde{g}_{\mu \nu}T^M_M
    \label{chapooy}
\end{eqnarray}
Comparing \eqref{chapoox}, \eqref{chapooy}  and taking trace over 
$\tilde{g}$,
\begin{eqnarray}
    \frac{1}{(D-2)}
    \frac{\nabla^2 \Omega^{D-2}}{\Omega^{D-2}}=R^{(d)} + \Omega^2( - T^\mu_\mu +   \frac{d}{D-2}T^M_M)\nonumber \\
\end{eqnarray}
As a result\cite{Maldacena:2000mw}, when we want to have positive curvature spacetime, i.e. $R^{(d)} > 0$ this implies,
\begin{eqnarray}
\boxed{
(D-d-2)T_\mu^\mu > d T_m^m.}\label{xchu}
\end{eqnarray}
Alternatively, a
similar statement is when  we want to have 
accelerated expansion 
$\frac{\ddot{a}}{a}\geq 0$
in warped compactification,
 the $D$-
dimensional Einstein field equations  allow a time-independent compactification
to a accelerating solution of dimension $d < D$ if the $D$-dimensional stress tensor 
violates the SEC.
%\begin{eqnarray}
%\boxed{
 %   R_{00}^{(D)}<0 \implies R_{MN}^{(D)}
  %  t^M t^N<0 }
%\end{eqnarray}
 This can be viewed easily in physical time coordinates of FRW metric. From eq. \eqref{x101}
we find out,
\begin{eqnarray}
    R_{00}^{(D)}=  -(d-1)\left(\dot{H}+ H^2 \right) +    \frac{\mathrm{
    \Omega}^{-(D-2)}}{D-2}\nabla^2\mathrm{\Omega}^{(D-2)}\nonumber\\
\end{eqnarray}
Multiplying both sides with $\Omega^{D-2}$ and
integrating over compact space leads to,
\begin{equation}
    (d-1)\frac{G_D}{G_d}\left(\dot{H}+ H^2 \right) =
-    \int d^n\tilde{y} \sqrt{\tilde{h}} \, \Omega^{(D-2)} R_{00}^{(D)}
\end{equation}
Here, $G_D$ and $G_d$ are Newton's constant in $D$ and $d$ dimensions respectively. 
We can clearly see to obtain the accelerating solution we need to violate (integrated version of) the $D$-dimensional SEC, i.e.
\begin{eqnarray}
    \int d^n\tilde{y} \sqrt{\tilde{h}} \, \Omega^{D-2} R_{00}^{(D)} < 0 \label{kasumi}
\end{eqnarray}
If the higher dimensional SEC was satisfied we would always have $\frac{\ddot{a}}{a}\leq 0$, i.e. non accelerating cosmology in the four dimensional external directions.
We should also note when we have accelerating FRW solution it automatically indicates-
   \begin{equation}
   R^{(4)}=3\left(\frac{\ddot{a}}{a}+(\frac{\dot{a}}{a})^2\right)>0, 
   \end{equation} positive curvature\footnote{Thanks to the referees for pointing it out}.
As the second term is clearly non-negative, by nature.
We will further prove in the next section that 
the accelerating backgrounds in  \eqref{x2}
should
 maintain the condition \eqref{kasumi} 
to satisfy the Raychaudhuri equation.
%\tableofcontents
\section{The Raychaudhuri equation and GMN no go theorem}

We
start to examine with background
 \eqref{x2}, if this
can satisfy the Raychaudhuri equation following \cite{bret}.
In other words, we would like to understand what are the constraints for the background \eqref{x2}
to maintain the Raychaudhuri equation. 
We first take the affine null vectors such as,
\begin{eqnarray}
  N^M=\frac{1}{\Omega^2}(1,0,0,0,\tilde{n}^m)
\end{eqnarray}
Here, $\tilde{n}_m$ is an affine unit $n$-dimensional spacelike vector with respect to metric $\tilde{h}_{mn}$, which means\cite{bret}-
\begin{eqnarray}
\tilde{h}_{mn}\tilde{n}^m\tilde{n}^n=1, \,\, \tilde{n}^m\tilde{\nabla}_m \tilde{n}^n=0 \label{kosherkosher}
\end{eqnarray}
The expansion parameter associated with $N_A$ is,
\begin{eqnarray}  \theta=N^m\partial_m[\ln(\Omega^{D-2}\sqrt{\tilde{h}}) ] + \frac{3H}{\Omega^2}\label{xchuuudada}
\end{eqnarray}
Here, $\sqrt{\tilde{h}}=[det(\tilde{h}_{mn})]^{\frac{1}{2}}$.
The    shear tensors have the following non-zero components.
\begin{eqnarray}
   && \sigma_{tt}= -\tilde{h}^{mn}\partial_n(\ln \Omega)N_m\label{a}\\
    &&\sigma_{tm}=\partial_m(\ln \Omega)\label{b}\\
    &&\sigma_{mn}=
-\Gamma^p_{mn}N_p-\frac{\theta}{D-2}\hat{h}_{mn}\label{c}\\
&&\sigma_{ij}=\Gamma_{ij}^M N_M
-\frac{\theta}{D-2}\hat{h}_{ij}\label{d}
\end{eqnarray}\\Here, $\Gamma^A_{BC}$ are the Christoffel Coefficients of \eqref{x2}. We have listed them in the appendix\cite{Bernardo:2022ony}. 
 Combining the equations \eqref{ava}, \eqref{xchuuudada} and \eqref{a}-\eqref{d}, we find out exactly as ref. \cite{bret},
\begin{eqnarray}
&&3(H^2+\dot{H}^2)=\tilde{R}_{mn}^{(n)}
\tilde{n}^m \tilde{n}^n +
A_{mn}(\Omega)\tilde{n}^m \tilde{n}^n\nonumber\\&&-{\Omega^4}R_{MN}^{(D)}N^M N^N \label{rooafza}
\end{eqnarray}
Here, $A_{mn}(\Omega)$ is defined as,
\begin{eqnarray}
    A_{mn}(\Omega)=(D-2)[\partial_m (\ln \Omega)
    \partial_n (\ln \Omega) -\nabla_m \partial_n (\ln\Omega)  ].\nonumber 
\end{eqnarray}
Rewriting \eqref{rooafza} as,
\begin{eqnarray}
3(H^2+\dot{H})
=\tilde{R}_{mn}^{(n)}\tilde{n}^m \tilde{n}^n +
A_{mn}(\Omega)\tilde{n}^m \tilde{n}^n\nonumber\\ 
-R_{00}^{(D)}-
R_{mn}^{(D)}\tilde{n}^m \tilde{n}^n\label{chapooshit}
\end{eqnarray} 
As we can see $R_{mn}^{(D)}$ is related to
$\tilde{R}_{mn}^{(n)}$
by conformal symmetry similar to \eqref{chapoox} which we can use to further simplify our calculation (using \eqref{bigchoppa}).
    \begin{eqnarray}
{ R}_{mn}^{(D)} = \tilde{R}_{mn}^{({n})}({\tilde{h}}) -
    \tilde{h}_{mn}\big((D-2)\partial_p\ln\Omega\partial^p\ln\Omega+\Box\ln\Omega\big)\nonumber\\
+(D-2)\big(\partial_m (\ln \Omega)
    \partial_n (\ln \Omega) -\nabla_m \partial_n (\ln\Omega)  \big),\nonumber\\ \label{abdulgaji}\end{eqnarray}
We can clearly see the last term in the 
    above equation \eqref{abdulgaji}
  exactly cancels $A_{mn}(\Omega)$ in eq. \eqref{chapooshit}. So we re write eq. \eqref{chapooshit},
\begin{eqnarray}
3(H^2+\dot{H})=-R_{00}^{(D)}
+\big(
(D-2) 
\partial_p\ln\Omega\partial^p\ln\Omega+\nonumber\\
\Box \ln\Omega\big)\tilde{h}_{mn}\tilde{n}^m
\tilde{n}^n
\label{chapooshit2}
\end{eqnarray}   
Simplifying further \eqref{chapooshit2} using the first equation of \eqref{kosherkosher}\footnote{We have also used a known mathematical identity $(D-2) \partial_p\ln\Omega\partial^p\ln\Omega+\Box \ln\Omega = \frac{\Box\Omega^{D-2}}{(D-2)\Omega^{D-2}}$, also appeared in eq.(33) of \cite{Maldacena:2000mw}} and  multiplying both sides by $\Omega^{{D-2}}$,
%\begin{eqnarray}
%3(H^2+\dot{H})=-R_{00}^{(D)}+
%\big((D-2)(\frac{\partial_p\Omega}{\Omega})^2+\nabla^p(
%\frac{\partial_p \Omega}{\Omega})
%\big)\nonumber 
%\end{eqnarray}
%Simplifying further,
\begin{eqnarray}
3(D-2)(H^2+\dot{H})\Omega^{D-2}=-(D-2)R_{00}^{(D)}\Omega^{D-2}+
{\Box \Omega^{D-2}
}
\nonumber \\
\end{eqnarray}
First, let us now perform an integral over the compact internal space, \\\begin{align}
&3(H^2+\dot{H}) \frac{G_D}{G_d}
= -\int d^n\tilde{y} \sqrt{\tilde{h}} \, \Omega^{D-2} R_{00}^{(D)}.
\label{UTD}  
\end{align}
Throughout, we have used the fact that the integral of the Laplacian of the warp factor over the compact manifold is zero as the warp factor is non-singular and the compact manifold has no boundary.
Let us first focus on -\\\\
A) The \textbf{dS solution},
where we know that $\dot{H}=0$.
As a result, the LHS is a positive definite quantity.\\\\
B) In the case of other  accelerating FRW
solutions,
\begin{eqnarray}
  \dot{H} + H^2=\frac{\ddot{a}}{a} >0
  \nonumber
\end{eqnarray}
and finally, LHS of \eqref{UTD} is also a positive definite quantity as well.
In both cases, we find out that
the LHS is positive.    
The
only way the RHS of \eqref{UTD} is a positive quantity if,
\begin{equation}
\int d^n\tilde{y} \sqrt{\tilde{h}} \, \Omega^{D-2} R_{00}^{(D)} < 0
\label{kat}
\end{equation}
this looks line an (integrated) constraint which indicates
towards (averaged) SEC violation in higher dimensions.
But this is the main essence of the GMN no-go theorem \eqref{kasumi}, i.e
one needs to violation SEC in the higher dimensions to obtain four dimensional 
accelerating FRW solution.
As we point out here the  Raychaudhuri equation results an integrated (averaged) SEC violation constraint. 
This follows entirely following a geometric identity, i.e., the
Raychaudhuri equation,
not having any presumption on the matter content of the theory.
In ref. \cite{bret},
the authors have argued
the constraints one obtains to solve this geometric identity
leads to much stronger condition than the existing no-go theorems such that the NEC must be violated at every point.
Therefore the NEC violation constrain put considerable restriction to construct any accelerating cosmology. 
The NEC is the weakest of the energy conditions in
a sense that
 a
violation of the NEC
implies a violation of the
other energy conditions, such as
weak, dominant, and strong
energy conditions.
The minimal coupling of NEC violating matter to Einstein gravity is likely inconsistent with string theory and black hole thermodynamics\cite{Chatterjee:2015uya}.
Also,
NEC violating theories
often display unsettling characteristics, namely superluminal propagation\cite{Dubovsky:2005xd},
unbounded negative Hamiltonians
\cite{Sawicki:2012pz}.
Therefore, if NEC violation is essential for the background \eqref{x2} to satisfy the Raychaudhuri equation, it could be problematic as NEC violation would cause the aforementioned problems. \\\\ But as we have investigated the problem carefully, we have realized from eq. \eqref{UTD} that 
it is important
to violate the
$D$ dimensional SEC (at least the integrated version\eqref{kat}) 
to satisfy the Raychaudhuri equation when backgrounds have accelerating FRW solutions in the external directions.
  This geometric identity \eqref{ava} does not impose any condition on the curvature of compact internal space and/or on the NEC for the background under consideration in \eqref{x2}. As a consequence violating NEC is not an essential condition for accelerating geometries.
Satisfying this geometric identity for the background \eqref{x2} leads to a constrain which indicates the
same conclusion
as GMN no-go theorem.\\\\ This can be understood from the seminal work of Jacobson. The main result of the GMN no-go theorem is dependent upon satisfying the Einstein equation. We basically take the trace over the higher dimensional Einstein equation in the external direction (see Section III).
 We do not explicitly perform such a step to find the relevant
constraints to satisfy the Raychaudhuri equation. But as pointed out by  works of Jacobson\cite{Jacobson:1995ab,Eling:2006aw},
%the Einstein equation may be derived from the entropy area along with the fundamental relation 
%$\delta Q = T dS $,
%connecting heat, entropy, and temperature.
% The heat flux $\delta Q$ is related to the energy-momentum tensor.
%On the other hand,
% the variation of the area is connected with the expansion of a bundle of null geodesics whose dynamics are governed by the null Raychaudhuri equation (see eq. (3) of \cite{Jacobson:1995ab}).
making the use of
null Raychaudhuri equation, along with the fundamental heat flow equation $\delta Q = T dS $
one can  obtain the Einstein equation as the equation of state.
%Therefore, when Raychaudhuri equation is solved along with this
%fundamental relation $\delta Q = T dS $,
%Einstein equation will automatically satisfy.
As a result, although we have not explicitly evoked the Einstein Equation in the derivation of section IV, it is already implied due to  Jacobson\cite{Jacobson:1995ab, Jacobson:2015hqa}. 
One of the main reason we reach a different conclusion compared to 
\cite{bret}, is when the authors reach 
\eqref{chapooshit} they do no relate 
$\tilde{R}_{mn}^{(n)}
$ with $ R_{mn}^{(D)}
$.
However, as we know these quantities are related by simple conformal transformation
\eqref{abdulgaji}. 
When we use it to derive the constrain, we find out that not only the contribution from spacelike  parts completely disappears but also the warp factor contribution can be nicely removed when we perform an integral over compact space. Consequently, we are left with an integrated constrain  involving  \textit{only} timelike directions of the Ricci tensor, i.e. $R_{00}$. This allows us to conclude that we need to violate the SEC to satisfy the Raychaudhuri equation and in the process we re-invent the main essence of GMN no-go theorem. 
We have not discussed the status of apparent horizon 
and antitrapped surfaces in this article\cite{bret}. In future  we would like to revisit these issues on time dependent background, because as pointed out by Townsend\cite{Townsend:2003fx,Townsend:2004zp} and Steinhardt\cite{Steinhardt:2008nk} to evade all the no-go theorems for dS compactification we might need to go beyond the time independent setup. In particular we would 
like to understand how such backgrounds can be realised as coherent state\cite{Bernardo:2021rul,Dvali:2018jhn} to bypass the Swampland conjectures\cite{Obied:2018sgi} and Trans-Planckian problems\cite{Bedroya:2019snp}. 
\begin{acknowledgments}
We thank F. Rost, 
L. 
 Aalsma, S. Ahmed,
H. Bernardo  and K. Dasgupta 
for discussions and comments on the draft.
\end{acknowledgments}
\appendix
\section{Christoffel Coefficients}
The list Christoffel Coefficients associated 
with \eqref{x2} are,
\begin{eqnarray}
&&  \Gamma^\rho_{\alpha m}=\partial_m (\ln \Omega) \delta^\rho_\alpha\nonumber \\
 && \Gamma_{mn}^\rho=0\nonumber\\
 && \Gamma^m_{\mu r}=0\nonumber\\
&& \Gamma_{\mu\nu}^m =\tilde{g}_{\mu\nu}\tilde{h}^{mn}\partial_n(\ln \Omega)\nonumber\\
 && \Gamma^m_{n r}=\tilde{\Gamma}^m_{n r}(\tilde{h})
 +\frac{1}{\Omega}(\delta_{r}^{m}
 \partial_{n}\Omega
 +
 \delta_{n}^{m}
 \partial_{r}\Omega
 -
 \tilde{h}_{nr}\partial^m\Omega)\nonumber
 \\
 && \Gamma^\rho_{\mu\nu}=\tilde{\Gamma}^\rho_{\mu\nu}(\tilde{g})\nonumber
\end{eqnarray}\\
  \section{Ricci Tensor}
  In any warped geometry represented with metric 
  \begin{eqnarray}
      ds^2=\tilde{h}(z)\tilde{ds}^2=
    \tilde{h}(z)  \tilde{G}_{MN}dz^Mdz^N\nonumber\\
  \end{eqnarray}
  The Ricci tensor for the full metric $R(G_{MN})$ is related to Ricci tensor of the metric 
  $R(\tilde{G}_{MN})$,
  \begin{eqnarray}
    &&  R(G_{MN})=R(\tilde{G}_{MN})-(D-2)\big(\nabla_M\nabla_N(\frac{\ln \tilde{h}}{2})
\nonumber\\
&&          -\nabla_M(\frac{\ln \tilde{h}}{2})\nabla_N(\frac{\ln \tilde{h}}{2})\big)  -\tilde{h}_{MN}\big(\nabla_P\nabla^P
(\frac{\ln \tilde{h}}{2})\nonumber\\&& +(D-2) \nabla_P(\frac{\ln \tilde{h}}{2}) \nabla^P(\frac{\ln \tilde{h}}{2}) 
\big)\label{bigchoppa}
  \end{eqnarray}
Block letter indices such as 
$M$ can be thought of direction $0,,,,D-1$  for the metric \eqref{xGG}.

\bibliography{apssamp}% Produces the bibliography via BibTeX.

\end{document}